\documentclass{aa}
\usepackage{graphicx}
\usepackage{epsfig,times}
\usepackage{color}
\usepackage{amssymb}
\usepackage{natbib}
\bibpunct{(}{)}{;}{a}{}{,} 

\begin{document}

\title{Supernova SN2006gy as a first ever Quark Nova?}

\author{Denis Leahy and Rachid Ouyed}

\institute{Department of Physics and Astronomy, University of Calgary, 
2500 University Drive NW, Calgary, Alberta, T2N 1N4 Canada\thanks{email:ouyed@phas.ucalgary.ca}}

\date{Received <date>; accepted <date> }

\authorrunning{Leahy\&Ouyed}

\titlerunning{SN2006gy as a first ever Quark-Nova?}

\abstract{
The most luminous Supernova SN2006gy (more than a 100 times brighter than a typical
supernova) has been a challenge to explain by standard models. For example,
 pair instability supernovae  which are luminous enough seem to have too slow a rise,
  and core collapse supernovae do not seem to be luminous enough. 
We present an alternative scenario involving a quark-nova  (an explosive
transition of the newly born neutron star to a quark star) in which a
second explosion (delayed) occurs inside the ejecta of a normal supernova.
 The reheated supernova ejecta can radiate at higher levels for longer periods
 of time primarily due to reduced adiabatic expansion losses, unlike
 the standard supernova case. 
 We find an encouraging match between the resulting lightcurve and that
 observed in the case of  SN2006gy  suggesting that we might
  have at hand the first ever signature of a quark-nova. Successful application of our model
   to SN2005gj and SN2005ap is also presented.
\keywords{stars: evolution --- stars: neutron --- supernovae: individual (SN2006gy) -- dense matter} 
}

\maketitle

\section{Introduction}

 Supernova (SN) SN2006gy is the most luminous supernova yet observed (more than 100
 times brighter and  significantly  longer-lasting  than a typical supernova); it 
  has so far challenged existing models. The fundamental question is
 how to power the observed lightcurve for so long (Smith et al. 2007; Ofek et al. 2007).
 Smith et al. (2007) rule out circumstellar medium (CSM) interaction based on the low observed X-ray flux and other  properties compared to known CSM powered SNe. On the other hand, 
  Ofek et al. (2007) argue that X-rays would be absorbed in the CSM so that the
  lack of X-rays does not rule out the CSM interaction mechanism. 
  Smith et al. (2007) instead
  favor  pair instability supernova (PISN) model  by energy argument; 22 $M_{\odot}$ of
$^{56}$Ni is needed to account for the peak luminosity. 
Scannapieco et al. (2005) considers PISNs for masses between 150$M_{\odot}$
 and 250$M_{\odot}$. PISN models brighter than $M_{\rm AB}\sim -21$ occur only for
  the most massive stars.  Langer et al. (2007)  explore
   the metallicity range  for PISN and conclude they can occur in the local universe.
    Despite the fact that PISN
 models of Scannapieco rose too slowly compared to SN2006gy no better alternative
 was available to Smith et al. (2007).  
  
 Nomoto et al. (2007) further consider PISN of 166 $M_{\odot}$ ejected
 mass and 15 $M_{\odot}$ ejected $^{56}$Ni.  They clearly  demonstrate that the lightcurve has
 too slow of a rise to be consistent with SN2006gy. 
  They can artificially fit the early parts ($\le 120$ days) 
 of the SN2006gy lightcurve by a PISN model with reduced ejected mass
  of about 50 $M_{\odot}$ (including the 15 $M_{\odot}$ of $^{56}$Ni), but  
   point out that such a low ejected mass is inconsistent with the 15 $M_{\odot}$ $^{56}$Ni mass.
  Umeda \& Nomoto (2007) as an alternative to PISN reconsider nucleosynthesis
 in core collapse explosions (initial mass less 100 $M_{\odot}$). Their main findings are that
 the $^{56}$Ni mass depend on initial stellar mass (which determines progenitor C$+$O core mass)
  and explosion energy (which determines core mass fraction converted to $^{56}$Ni).
   The maximum $^{56}$Ni mass (of $\sim 13M_{\odot}$) 
   was obtained for the most massive star with $M_{\rm CO}\sim
   43 M_{\odot}$ and explosion energy of $\sim 2\times 10^{53}$ ergs which they
   note to be unrealistically large.

   Two recent papers make use of the idea of shock
    energy being deposited in an extended envelope
     to minimize adiabatic losses so that the light curve
     of SN2006gy can be powered by shock energy. 
Woosley,  Blinnikov, \&  Heger (2007) consider 
a pulsational pair-instability supernova, which  leads to a second ejection.
The interaction between the second and the first ejection powers the
 light curve, in contrast to a normal PISN which is powered mainly
  by a large $^{56}$Ni mass as discussed above.
This gives rise to a light curve bearing some similarities to SN2006gy. 
   Smith\&McCray (2007) give a general argument that  the light curve can be produced
    by shock propagating in an envelope with an initial radius of order
     of 60 AU, which avoids adiabatic expansion losses.

It has long been thought that the
    center of neutron stars may be dense enough that nuclear boundaries
     dissolve and a phase transition to  matter 
     made of up and down quarks occurs (Itoh 1970; Bodmer 1971).
      It was then conjectured by Witten (1984) that the
      addition of the strange quark to the mixture would lead to a
      true ground state of strongly interacting matter at zero pressure
       making the existence of quark stars an intriguing possibility (Alcock et al. 1986).
        Neutron star cooling studies show no strong incompatibility with
standard neutron star models (e.g. Page 2004). However quark stars are not ruled
out either: mass-radius studies still allow quark star
 equations of state (e.g. Leahy 2004).       The transition from neutron star to strange star associated with SNe  has been suggested  previously  (e.g. Horvath \& Benvenuto 1988;
Drago et al. 2007 and references therein).  In addition
 a two-bang scenario was proposed in the context of SN1987A to
explain  the delayed neutrinos, where
 the delay is due to the collapse of the neutron star
into a black hole or strange star (De Rujula 1987).

   We propose a  model based on additional energy input into the supernova
    ejecta:  (i) The explosion occurs inside an extended
expanding envelope; (ii) The delayed explosion is due to conversion
 of neutron star (NS) to a quark star (QS).
 No one has used the conversion from NS to QS to explain extremely bright SNe,
nor has anyone used the crucial idea of delayed explosion.
  Benvenuto\&Horvath (1989)  explored the idea
   of conversion energy release to power SN1987A, which
is a regular SNe. However, they do not calculate any lightcurves,
 do not consider explosive conversion, nor do they make use of the 
  conversion delay.
   In our work here,  the additional energy input into the supernova
    ejecta is a consequence of an explosive conversion of a neutron star to
    a quark star namely, a  Quark-Nova (QN). 
      The new ideas  here are: (i) the explosive
conversion in the QN; (ii)  the resulting re-energization of the SN ejecta.
 The QN  energy input is delayed
    from the original core collapse explosion, allowing for re-energization of
    the SN ejecta at larger radius. As we show in this paper this allows for
    more luminous and long-lasting explosion since 
    much of the radiation is emitted rather than being lost to adiabatic expansion.
     We first discuss the QN process.

In the QN picture (Ouyed et al. 2002;  Ker\"anen\&Ouyed 2003; Ker\"anen et al. 2005) 
 it was shown that detonation rather than deflagration occurs, so that the converted core contracts and separates from the un-converted crust of the neutron star.
 Specifically,  the core of a neutron star, that undergoes the phase transition to the quark phase, 
shrinks in a spherically symmetric fashion to a stable, more compact quark  
matter configuration faster than the overlaying material (the neutron-rich
hadronic envelope) can respond.
The resulting quark star initial temperature is of the order of 10-20 MeV
  since the collapse is adiabatic  rather than isothermal (Ker\"anen, Ouyed, \& Jaikumar  2005).
  The energy released during the QN explosion can be as high as  $E_{\rm QN}\sim 10^{53}$ erg  and 
 involves baryon to quark conversion energy and gravitational energy
release due to contraction. 
Unlike a core collapse supernova, a large
fraction of the energy of a QN after the collapse is released in photons.  This is due to
 to unique properties of quark matter in the superconducting color-flavor-locked (CFL) phase
  (Rajagopal\&Wilczek 2001;  for more recent studies on the feasibility
   of the CFL phase and its properties see Pagliara\&Schaffner-Bielich 2007 and Alford et al. 2007).  
    As shown in Vogt et al. (2004) and  Ouyed et al. (2005), the CFL phase 
  favors photon emissions to standard neutrino ones. 
   The time delay between the SN and the QN is controlled mainly 
 by spin-down and the increase in core density  of the neutron star (Yasutake et al. 2005; Staff et al. 2006),  and secondly  by the weak conversion between quark flavors (e.g. Bombaci et al. 2004).
   
  The QN ejecta ($< 0.1 M_{\odot}$; see Ker\"anen et al. 2005
 and Jaikumar et al. 2007), 
 which is the left-over crust of the parent neutron star,
is initially in the shape of a shell and is imparted with energy from the QN explosion.
One can show that it is expanding relativistically with Lorentz
factor of a few  (Ouyed et al. 2007).  Inside the SN ejecta the QN ejecta
rapidly sweeps up enough mass to become sub-relativistic. In simple
terms this sets up a second blast wave propagating outward and 
reheating the SN ejecta.  This second blast wave causes reheating of the ejecta
 at larger radii, thus adiabatic losses occur
  on much longer timescale than for the initial SN explosion. This
  is the key to the long duration and high brightness of the radiation
  from the second (QN) shock. Simply put, in the normal SN the shock
  radiation is lost to adiabatic expansion before it can diffuse out
  while for the delayed shock case much of the radiation diffuses out before
   adiabatic losses dominate.

This paper is presented as follows: Section 2 describes the SN phase.
Section 3 describes the effects of the second explosion on the SN lightcurve
 with application to SN2006gy. Section 4  deals with SN2005gj before briefly
concluding in Section 5.

\section{The SN phase: {\it The first shock}}\label{sec:SN}

Let us assume that a SN has exploded and processed its ejecta
by explosive burning. In our simplified model the energy from the shock
is deposited instantly heating the ejecta to an initial temperature $T_{\rm SN,0}$. 
This initial state is that of an expanding ejecta with a central region of 
  $^{56}$Ni. The ejecta is uniformly expanding in time 
   (i.e. the velocity is linear with radius at any fixed time) with the
   outer radius of the ejecta given by  $R=R_0 + v_{\rm SN} t$
 where $R_0$ is the size of the progenitor star and $v_{SN}$ is
 the speed of the shocked SN material. 
 
  Due to outward diffusion of photons the atmosphere is moving inward in mass coordinates,
   slowly at first but faster as the density decreases in time. The ejecta
   interior to the atmosphere we refer to as the core.
 We will assume that the thermal energy in the exposed mass in the atmosphere 
  (as the cooling front  creeps inward) is instantly radiated. The interplay between 
  uniform expansion and radiation diffusion defines the evolution of the photosphere as
  \begin{equation}
 R_{\rm phot.} (t) = R_0 + v_{\rm SN} t - D(t)\ ,
 \end{equation}
 where $D(t)$ is the diffusion length  
 \begin{equation}
 D(t)^2 = D_0^2 + \frac{c}{n_{\rm ejec.}\sigma_{\rm Th}} t\ ,
 \end{equation}
 where $c$ is the speed of light, $\sigma_{\rm Th}$ is the Thompson cross-section, and
  $n_{\rm ejec.}= N_{\rm ejec.}/V_{\rm ejec.}$ is
 the particle density in the ejecta.
  For an ejecta of mass  $M_{\rm ejec.}$ and mean molecular weight $\mu $,  the total
  number of particles is
   $N_{\rm ejec.}= (M_{\rm ejec.}/\mu m_{\rm H})$ where $V_{\rm ejec.} = (4\pi/3) (R_0+v_{\rm SN}t)^3$
    is the volume extended by the ejecta and $m_{\rm H}$ is the Hydrogen atomic mass. 
We define $D_0$ as the initial diffusion lengthscale
 by setting $n_{\rm ejec., 0}\sigma_{\rm Th} D_0\simeq 1$
  where $n_{\rm ejec.,0}= N_{\rm ejec.}/V_{\rm ejec.,0}$ with the initial
  volume $V_{\rm ejec., 0} = (4\pi/3) R_0^3$.
  
The corresponding luminosity is
\begin{equation}
\label{eq:luminosity}
L_{\rm SN}(t)  =  c_{\rm v} \Delta T_{\rm core} n_{\rm ejec.} 4\pi R_{\rm phot.}(t)^2 \frac{d D(t)}{d t}\ ,
\end{equation}  
 where the  specific heat is $c_{\rm v}\sim (3/2) k_{\rm B}$
  and $\Delta T_{\rm core}\sim T_{\rm core}$ since the atmosphere cools instantly (i.e.
  cooling time is much less than the diffusion timescale); $k_{\rm B}$
   is the Boltzmann constant. The
  rate of mass flux from the core to the photosphere is determined by the velocity $d D(t)/dt$.
  
  Ignoring input from radioactive decay, 
adiabatic expansion of the core leads to
  \begin{equation}
  T_{\rm core} = T_{\rm SN,0} \frac{R_{0}^2}{(R_0+ v_{\rm SN} t)^2}\ ,
  \end{equation}
  which is used when computing the SN luminosity.
  The effect of radioactivity 
  consists of heat added to the core from $^{56}$Ni and $^{56}$Co decay
 keeping the core temperature high for weeks to months even in the
presence of adiabatic expansion losses.  For standard SNe both type I 
(e.g. Sutherland\&Wheeler 1984) and most type II (e.g. Suntzeff et al. 1992)
the late time lightcurve is dominated by radioactivity. However this does not seem
to be the case for SN2006gy  since the radioactivity lightcurve
peaks much later than the observed peak (see Figure 9
in Nomoto 2007).


\section{The Quark Nova phase: {\it The second shock} }\label{sec:QN}

 The QN goes off at $t_{\rm QN}$ after the SN explosion. The QN shock 
  propagating at speed $v_{\rm QN}$
  reaches the outer edge of the SN ejecta (becomes visible to the observer)
  at distance $R_{\rm QN}$ and time $t_{\rm QN}+ t_{\rm prop.}$ where $t_{\rm prop.}= R_{\rm QN}/v_{\rm QN}$ is the propagation time delay for the QN shock to reach the edge. 
 That is, the ejecta is first fully reshocked at a  radius
 $R_{\rm QN} = R_0+v_{\rm SN} (t_{\rm QN}+R_{\rm QN}/v_{\rm QN})$ heating up the SN material 
 to a new temperature $T_{\rm QN, 0}$.  The evolution of the  new
 photosphere is then
  \begin{equation}
 R_{\rm phot.} (t) \simeq  R_{\rm QN} + v_{\rm QN} t - D_{\rm QN}(t)\ ,
 \end{equation}
 where $D_{\rm QN}(t)$ is the diffusion length with parameters
 reset at $t_{\rm QN}+R_{\rm QN}/v_{\rm QN}$.
  Again, ignoring input from radioactive decay in the core, adiabatic expansion
   gives
  \begin{equation}
  T_{\rm core} = T_{\rm QN,0} \frac{R_{QN}^2}{(R_{\rm QN}+ v_{\rm QN} t)^2}\  .
  \end{equation}
  In a normal SN,  adiabatic expansion rapidly cools the ejecta
 to 3000K. Here the SN ejecta within the photosphere stays hot after the QN shock
 for a long time. A simple estimate using  eq.(6) yields
 70 years before cooling to $3000$ K. 
  For example, including Bremsstrahlung cooling,  for the first 10 days
 the ejecta outside the photosphere cools rapidly. However, it represents
 only a tiny fraction of the mass of the shocked SN ejecta.
 The bulk of the shocked SN ejecta, which moves outside the photosphere
after 10 days, expands to low density quickly enough that the cooling can be neglected
 (due to the $n^2$ dependence of cooling).
  
\begin{figure}
\centerline{\includegraphics[width=0.5\textwidth,angle=0]{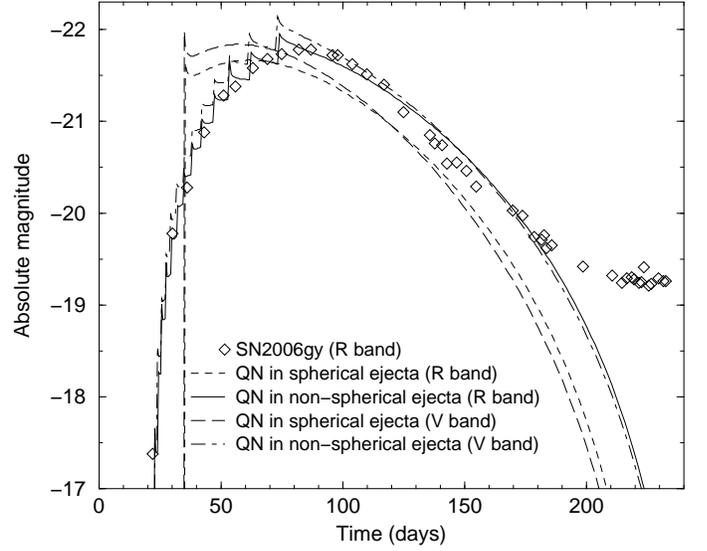}} 
\caption{\label{fig:sn2006gy}
 Comparison of the observed absolute $R$-band light curve of SN2006gy and the $R$-band
  and $V$-band light curves derived
 from our model. 
 The dashed (long-dash) line shows the derived $R$-band ($V$-band) light curve for  
 a QN explosion inside perfectly spherical SN ejecta.  The solid (dash-dot) line
  shows the derived $R$-band ($V$-band) light curve
  for a QN explosion inside a non-spherically expanding SN ejecta.
 The SN parameters are: Explosion at $t=0$, $M_{\rm eje.}= 60 M_{\odot}$, $R_0=10 R_{\odot}$, and the  QN parameters  are $t_{\rm QN}=15$ days,  $v_{\rm QN} = 6000$ km s$^{-1}$, and  $T_{\rm QN, 0}= 0.4$ MeV.  For the spherical case $v_{\rm SN}= 3400$ km s$^{-1}$ while
 for the non-spherical case $2000\ {\rm km s}^{-1} < v_{\rm SN} < 4800\ {\rm km s}^{-1}$.
 The spikes in the derived light curves are due to pieces of the SN ejecta
     being lit up by the QN shock at different times, which would be 
     smoothed out if the distribution of velocities were continuous.
     The Smith et al. (2007) data was plotted with the first data point (which is an upper limit) 
     at $t= 22$ days in order to match our model with the overall rise.
 }
\end{figure}

  In the case where the QN impacts into a perfectly spherical SN ejecta,
   the calculated luminosity has a sharp rise when the QN shock reaches the outer edge
   of the SN ejecta.   
   This is followed by fairly flat period before a smooth
   decline due to the photosphere moving inwards ($R_{\rm phot.}$ decreasing).
    An example is shown  in Figure \ref{fig:sn2006gy} by the
    dash and long-dash lines (for $R$- and $V$-band, respectively) compared to the data  from SN2006gy (Smith et al. 2007). 
    The model corresponds to a SN explosion at $t=0$  with 
   $M_{\rm eje.}= 60 M_{\odot}$, $R_0=10 R_{\odot}$, $v_{\rm SN}= 3400$ km s$^{-1}$.
    The QN explosion occurs at $t_{\rm QN}=15$ days with velocity $v_{\rm QN}=
    6000$ km s$^{-1}$. The above velocities were based on 
     Smith et al. (2007) who find extended wings in $H_{\alpha}$ of $\sim \pm 6000$ km s$^{-1}$
     (our choice of $v_{\rm QN}$).
In addition there is a blueshifted $H_{\alpha}$ absorption up to $\sim 4000$ km s$^{-1}$, which
  could be a signature of the first shock on which we base our choice of  $v_{\rm SN}$.
   The total thermal energy
    deposited by the QN shock in the SN ejecta to reheat it to $T_{\rm QN, 0}\sim 0.4$ MeV
     is of the order $\sim 3\times 10^{52}$ erg which consistent with QN explosion
     energetics.
     The sharp rise in the model occurs at $(t_{\rm QN}+t_{\rm prop.})= (15+19.6)=34.6$ days. 
    No attempt was made to fit this model to the data due to the
    sharp rise in the model and the importance of asphericity on the
    lightcurve (see below).

\subsection{Effect of asphericity on the lightcurve}

As noted above, the model curve has a sharp turn on in the case of a QN explosion
 into a spherically expanding SN ejecta. The SN is likely to be asymmetric 
 primarily due to variation in expansion velocity $v_{\rm SN}$. We account for this
  by extending our model to take into account a range of $v_{\rm SN}$. 
  The main result
   is varying radius ($R_{\rm QN}$) and time  when the QN shock reaches the outer edge of the SN ejecta. That is, 
   \begin{equation}
   R_{\rm QN}(v_{\rm SN}) = \frac{R_0+ v_{\rm SN} t_{\rm QN}}{1-\frac{v_{\rm SN}}{v_{\rm QN}}}\ ,
   \end{equation}  
   leading to a time delay for different parts of the ejecta of
     $t_{\rm prop.}(v_{\rm SN}) = R_{\rm QN}(v_{\rm SN})/v_{\rm QN}$.
   We note that $v_{\rm QN} > v_{\rm SN}$ in order for the second shock to occur.
   If the range of velocities in the SN ejecta extends to lower values, the delay is less
      between the QN and the initial rise in the lightcurve.   
          
   The resulting light curve is a superposition of light curves from different
   parts of the reshocked SN shell, with different rise times,  different peaks, and
   different shapes. The corresponding light curve is shown in Figure \ref{fig:sn2006gy} (solid
    and dash-dot lines for $R$- and $V$-band, respectively) and 
    corresponds to an SN explosion at $t=0$ with 
   $M_{\rm eje.}= 60 M_{\odot}$, $R_0=10 R_{\odot}$, $t_{\rm QN}=15$ days,  $v_{\rm QN} = 6000$ km s$^{-1}$, and  $T_{\rm QN, 0}= 0.4$ MeV. 
   The lightcurve was computed by averaging over 13 equal solid angle segments
   of a sphere with different velocities linearly spaced between the minimum and maximum
    values: $2000\ {\rm km\ s}^{-1} < v_{\rm SN} < 4800\ {\rm km\ s}^{-1}$.
     The lightcurve first turns on  when the slowest ejecta ($v_{\rm SN, min}=2000$ km s$^{-1}$) is fully 
     reshocked at $t_{\rm QN}+t_{\rm prop.}(v_{\rm SN, min})= (15+7.5)=22.5$ days.
     The Smith et al. (2007) data was plotted with the first data point (an upper limit) 
     at $t= 22$ days in order to match our model with the overall rise. 
     The spikes in the lightcurve (dashed line) are due to pieces of the SN ejecta
     being lit up by the QN shock at different times, which would be 
     smoothed out if the distribution of velocities were continuous.
   The SN material at lower velocities experiences the QN shock earlier resulting
    in larger adiabatic losses and lower peak brightness. 
    We note that the first shock (namely the SN proper) 
     is too faint to be seen due to the large distance to SN2006gy. Even when
    we add $4 M_{\odot}$ of $^{56}$Ni to the first SN (this is the maximum $^{56}$Ni 
    produced for a $60M_{\odot}$ progenitor; Nomoto et al. 2007) we estimate a magnitude
    $M_{\rm R}\simeq -18.5$ at 22 days which is only slightly above the upper limit for detection.
     This may indicate that the SN produced less $^{56}$Ni than the maximum expected.

  \subsection{The plateau beyond 200 days}

  We first note that the maximum  $4 M_{\odot}$ of $^{56}$Ni from the SN 
   cannot power the   late time plateau at $M_{\rm R}\sim -19$.  
   Nomoto et al. (2007) points
   out that for core collapse explosions most of the C$+$O core that is exposed to a radiative shock with $T> 5\times 10^{9}$ K   is converted to $^{56}$Ni. 
  An interesting aspect of our model is that the second shock due to the QN is
  a hot radiative shock and can convert much of the C$+$O (and Silicon and Magnesium) 
  in the SN ejecta to $^{56}$Ni.
   We estimate the initially ejected C$+$O mass for the  $60M_{\odot}$ model to be
    $\sim 25M_{\odot}$ (see Table 2 in Umeda\&Nomoto 2007).  The first shock
     converts the C$+$O to successive layers of $^{56}$Ni, $^{28}$Si, and $^{16}$O/$^{24}$Mg (see
     Figure 5 of Umeda\&Nomoto 2007). We suggest that
     the QN explosion might convert enough of the  $^{28}$Si, and $^{16}$O/$^{24}$Mg to $^{56}$Ni 
     to power the plateau. These are all zero neutron excess
     nuclei and  as discussed by Umeda\&Nomoto (2007) shock nucleosynthesis at high
      temperatures in zero neutron excess matter primarily produces nearly pure $^{56}$Ni.
     Finally, let us note that the QN explosion can provide very high
      explosion energies (up to $\sim 10^{53}$ ergs) which further favors  nucleosynthesis of $^{56}$Ni.

\begin{figure}
\centerline{\includegraphics[width=0.5\textwidth,angle=0]{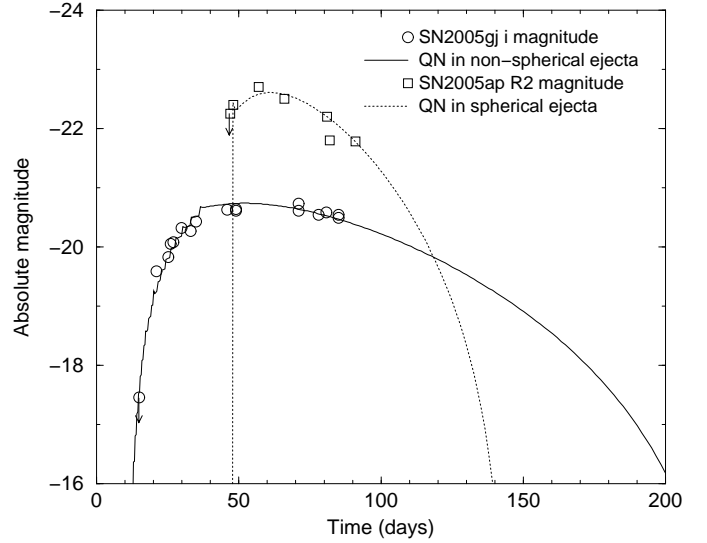}} 
\caption{\label{fig:sn2005gj}
 Comparison of the absolute $i$-band light curve of SN2005gj and $R2$-band light curve
  of SN2005ap with those derived  from our model.  For SN2005gj, the model is calculated  with a QN
  delay of 10 days after the SN,  and a range in SN
  ejecta speeds of $750\ {\rm km s}^{-1} < v_{\rm SN} < 4100\ {\rm km s}^{-1}$.
  For SN2005ap, the model is calculated  with a QN
  delay of 40 days after the SN,  and spherical SN
  ejecta with speed of $v_{\rm SN} = 4000\ {\rm km s}^{-1}$.
   For both models, all other QN and SN parameters 
   were kept the same as for the SN2006gy model. 
   In order to match our models with the rises in the data, 
     the SN2005gj data  (Aldering et al. 2006) was plotted with the first data point (which is an upper limit)  at $t= 15$ days, while the SN2005ap data 
      (sampled from Quimby et al. 2007) was plotted with with their $t=0$ point  at $t=57$ days.
 }
\end{figure}

 The late-time  lightcurve may be an
important way to differentiate between models.
Late-time observations of SN2006gy  (Smith et al. 2008) show that the
  observed luminosity at 400 days is  consistent with  $\sim 2.5M_{\odot}$
  of $^{56}$Ni, which is   too low to be consistent with the 
  $^{56}$Ni mass required to power the peak ($\sim 20 M_{\odot}$).
In effect the late-time light curve rules out $^{56}$Ni higher than $2.5M_{\odot}$. 
 Calculations of the total mass of  $^{56}$Ni in our model are very sensitive to
 the density in the ejecta, and cannot be reliably estimated with our
simplified model. This is left as future work. 
  
 The late-time lightcurve could alternately be re-emission of
  the SN peak light curve light by dust.  We favor this idea 
 since the late-time light curve  is detected in K band but not in R band (see their figure 2
  of Smith et al. 2008).    Specifically, Smith et al. (2008) argue that
   the dust is in a shell   of $\sim 10M_{\odot}$ ejected 
   1500 yrs prior to the SN.  Since the late-time light curve is  smoothly decreasing
   with time, rather than a burst,  we instead propose   that the dust is  ambient in the interstellar
   medium within a few pc of  SN2006gy.  From the late-time luminosity,
     using a gas-to-dust ratio of 100, we estimate a local gas number density
      of $\sim 20$ cm$^{-3}$ is enough to explain the late-time tail.
       If the late-time light is due to dust,  the relative brightness in K and H band would not measure the temperature
 evolution of the ejecta but rather measure the temperature of the dust.

\section{Supernovae SN2005gj and SN2005ap}

The light curve of SN2005gj is the second brightest SN ever observed showing
 similarities to SN2006gy.   Its light curve rose more quickly and to a 
higher peak luminosity than typical SNe, and declined much more slowly (Aldering et al. 2006).
 They are both classified as hybrid (i.e. a mixture of Type Ia and Type IIn spectra).
While it has been argued that its brightness might be a consequence of
 a strong interaction between the SN ejecta and the CSM, no 
X-ray (Immler et al. 2005) and radio (Soderberg \& Frail 2005) 
 have been detected. Applying our model to this candidate shows
  encouraging results as can be seen from Figure \ref{fig:sn2005gj} where the 
   $i$-band lightcurve from our model is compared to the observed one.
  We assumed that SN2005gj progenitor is similar to that of SN2006gy
   ($M_{\rm eje.}= 60 M_{\odot}$, and $R_0=10 R_{\odot}$) and that the QN
   features are also the same ($v_{\rm QN} = 6000$ km s$^{-1}$ and  $T_{\rm QN, 0}= 0.4$ MeV).
   The fit was obtained for $t_{\rm QN}=10$ days (i.e. the neutron star
   turned into a quark star sooner than in the SN2006gy) and by taking
    a slightly different range in ejecta speed,  $750\ {\rm km s}^{-1} < v_{\rm SN} < 4100\ {\rm km s}^{-1}$
     (probably due to small differences in SN progenitor or environment).   
Further monitoring of the SN2005gj in the $i$-band should help  distinguish
     between our model and those proposed in the context of CSM interaction 
     (e.g. Chugai\&Yungelson 2004; see also Figure 5 in Aldering et al. 2006).
     
  SN2005ap has just come to our attention  as possibly being
   the brightest SN supplanting SN2006gy (Quimby et al. 2007). Contrary to SN2006gy
    and SN2005gj  this candidate shows a rapid decline in its light curve.
     As a further test of our model, we apply it to this recently discovered SN.
     The spectrum of SN2005ap shows velocities greater than $\sim  23000$ km s$^{-1}$, 
        much higher than SN2006gy and SN2005gj, indicating very high
        QN shock velocity. Assuming a similar progenitor  as for the other
         candidates, we obtain a remarkably good fit to the light curve for a QN delay of $t_{\rm QN}=40$ days  and $v_{\rm QN}=25000$ km s$^{-1}$ (see Figure \ref{fig:sn2005gj}). In this case  a spherical SN ejecta ($< 15\%$
         asphericity, based on our models)  works well.
           In the dual explosion picture we present here, the longer
     the delay the lower the density of the inner edge of the SN ejecta when it is
      shocked by the QN ejecta. Thus we expect a higher QN shock
      velocity for longer QN delays, which seems to the case for SN2005ap.

\section{Discussion and conclusion}\label{sec:conclusion}

We have applied our model to the three most luminous SNe: SN2006gy, SN2005ap, and SN2005gj.
 The difference in parameters are the range in $v_{\rm SN}$ and the time delay
 $t_{\rm QN}$. One naturally expects variation in $v_{\rm SN}$. For $t_{\rm QN}$,
 the derived values range from 10 to 40 days, much longer than the dynamical
timescale of a compact object. However, the time delay between the SN and the QN is controlled 
 by spin-down and the increase in core density  of the neutron star (Yasutake et al. 2005; Staff et al. 2006),  and secondly  by the weak conversion between quark flavors (e.g. Bombaci et al. 2004).
 The core density of the neutron star first needs to reach deconfinement density
 (i.e.  conversion from hadrons to up and down quarks). Then weak conversion processes
 convert the (u,d) core to  strange quark matter (u,d,s).
 The spin-down delay to deconfinement density can range from less than one day to $\sim 1000$ years; the subsequent weak conversion delay
 is currently unknown. Our three derived total delay times were 10, 15 and 40 days. Since the total delay is the sum of a universal weak delay plus a variable spin-down delay, we constrain the weak delay to be less than 10 days. 
  
  Yasutake et al. (2005) and Staff et al. (2006) have determined that
 the evolutionary transition from rapidly rotating neutron stars to quark stars
 due to spin-down can lead to an event rate of $10^{-4}$-$10^{-6}$ per year
 per galaxy. Similar rates were derived from studies of QNe contributions
 to r-process material in the Galaxy by Jaikumar et al. (2007) who estimated that 1
 out every 1000 neutron stars might have undergone a QN.  Since the Galaxy
 likely contains about $10^{8}$ neutron stars this suggests an average QN rate of
 $10^{-5}$ per year per galaxy.  Interestingly, the fraction of SN progenitors
 with mass greater than $60M_{\odot}$ can be estimated
  as $\sim 5\times 10^{-3}$, using 
 the Scalo (1986) initial mass function for $M> 8 M_{\odot}$.
 Using a SN rate of  $\sim 10^{-2}$ per year per galaxy,
 we get $\sim 5\times 10^{-5}$ per year per  galaxy for the explosion rate of 
  massive star ($> 60 M_{\odot}$). This is, within uncertainties,  
  the same as the QN rate.

 Our model suggests that the lightcurve of SN2006gy is mainly due
to shock radiation from  a delayed
 explosion inside  an expanding SN ejecta of mass of $60M_{\odot}$. 
  To obtain the necessary peak luminosity for SN2006gy the second shock must reheat
  the SN ejecta to $\sim 0.4$ MeV and the reheating must occur
  at a large radius to minimize adiabatic expansion losses. The required energy for
   the reheating by the second shock is characteristic
  of a typical QN explosion.  The $\sim 15$ days delay time
   is derived by fitting the light curve. 
     In principle, the conversion delay time is not well constrained 
    by theory since: (i) the spin-down delay depends on unknown initial
    spin-period and mass; (ii) the conversion process
is very complex involving more than just weak processes.  
 In fact, fitting the light curves of extreme SNe
  may give  a means of inferring the microphysics of neutron to quark transition.
      To summarize, larger luminosities are obtained for reduced
       adiabatic losses which depend on the radius at which the QN
      shock breaks out of the ejecta. This can occur if
        the QN delay is long and the QN shock moves rapidly through the ejecta, or  if  the delay
       is short and the QN shock moves slowly through  the ejecta so  it takes a long time to break out. 
  
  Furthermore, if the SN ejecta density is high enough (i.e. $t_{\rm QN}$
  is small), the high temperature of the QN shock can process
   $^{12}$C, $^{16}$O, $^{28}$Si and $^{24}$Mg
  into $^{56}$Ni. This might provide late time emission which could explain the plateau beyond 200
  days for SN2006gy, which also may occur for SN2005gj (although we favor dust emission
  as the driver of the late-time light curve; see \S 3.2). 
  In contrast, we do not expect much extra $^{56}$Ni production
   for SN2005ap given its long QN delay. Finally, we mention that the neutron-rich QN ejecta
   is converted to r-process elements beyond $A=130$  which may be
   visible in the late time spectra of SN2006gy  -- these nuclei
and the associated observable $\gamma$-ray flux is tabulated in Jaikumar et al.
 (2007).   Specifically,  the photon flux from $\gamma$-decay of certain heavy r-process nuclei can
act as tags of the QN , differentiating them from
PISN (due to the lack of neutron excess) or core-collapse (lower
     neutron excess than the QN ejecta) alternatives.  
    Finally, we note that the QN explosion provides enough energy (up to $10^{53}$ ergs)
to power SN2006gy, while  the  PPISN model (Woosley et al. 2007) second explosion 
 provides $6\times 10^{50}$ ergs which was artificially increased by a factor
 of 4 to give the SN2006gy peak (see Figure 3 of Woosley et al. 2007).
 In addition, the time delay between the two explosions in each model
 differs: 10-40 days in the QN model versus $\sim 7$ years in the PPISN model.
   
  If QNe do indeed occur in the universe, as this work 
 seems to indicate, the consequences to astrophysics
  in general and to high energy astrophysics in particular (e.g. Niebergal et al.  2006; 
   Staff et al.  2007; Ouyed et al. 2007) could be tremendous.


\begin{acknowledgements}
We thank P. Jaikumar, J. E Staff,  and C. Foellmi for comments on the paper.
This research is supported by grants from the Natural Science and
Engineering Research Council of Canada (NSERC).
\end{acknowledgements}


\clearpage

\end{document}